\def \s{~\rm{s}}
\def \km{~\rm{km}}

\def \AU{~\rm{AU}}
\def \erg{~\rm{erg}}

\def \yr{~\rm{yr}}

\documentclass[12pt,preprint]{aastex}
\usepackage{natbib}

\begin{document}
\title{STEADY TWIN-JETS ORIENTATION: IMPLICATIONS FOR THEIR FORMATION MECHANISM}

\author{Noam Soker\altaffilmark{1} \& Liron Mcley\altaffilmark{1}}

\altaffiltext{1}{Dept. of Physics, Technion, Haifa 32000, Israel;
soker@physics.technion.ac.il; lironmc@tx.technion.ac.il}

\begin{abstract}
We compare the structures of the jets of the pre-planetary nebulae (PN) CRL618 and the young stellar object (YSO) NGC~1333~IRAS~4A2
and propose that in both cases the jets are launched near periastron passages of a highly eccentric binary system.
The pre-PN CRL618 has two `twin-jets' on each side, where by `twin-jets' we refer to a structure where one side is composed of
two very close and narrow jets that were launched at the same time. We analyze the position-velocity diagram of NGC~1333~IRAS~4A2,
and find that it also has the twin-jets structure. In both systems the orientation of the two twin-jets does not change with time.
By comparing these two seemingly different objects, we speculate that the constant relative direction of the two twin-jets is
fixed by the direction of a highly eccentric orbit of a binary star.
For example, a double-arm spiral structure in the accretion disk induced by the companion
 might lead to the launching of the twin-jets.
We predict the presence of a low-mass stelar companion in CRL618 that accretes mass and launches the jets,
and a substellar (a planet of a brown dwarf) companion to the YSO NGC~1333~IRAS~4A2 that perturbed the accretion disk.
In both cases the orbit has a high eccentricity.
\end{abstract}


\section{INTRODUCTION}
\label{sec:intro}

Many planetary nebula (PNe) are thought to be shaped by jets that are launched from a binary companion
(\citealt{SokerLivio1994, SahaiTrauger1998}; see review by \citealt{BalickFrank2002}), in particular
two opposite dense clumps/bullets (termed also ansae or FLIERS) are though to be formed by jets \citep{Soker1990}.
Understanding the properties of the jets and the conditions for their formation is relevant
to other astrophysical systems as well, such as young stellar objects (YSOs) that share some similar properties with
some PNe \citep{LeeSahai2004}.
Examples include the well collimated jets from Hen~2-90 that are composed from chain of bullets that are similar to
jets from some YSOs \citep{SahaiNyman2000},
and the similar morphology of the PN KjPn~8 (for a recent study of this PN see \citealt{BoumisMeaburn2013})
and the bipolar structure of the recently discovered YSO Ou4  \citep{Acker2012, Corradietal2013}.
In the present study we focus on the pre-PN CRL618 recently analyzed by \cite{Balicketal2013}, and
compare the two opposite \emph{`twin-jets'} with those in the YSO NGC~1333~IRAS~4A2 (hereafter IRAS~4A2).
By \emph{twin-jets} we refer to two narrow and close jets (or bullets) that are launched at the same time.
\cite{Leeetal2003} already compared CRL618 with the YSO HH~240/241.
\cite{Leeetal2003} and \cite{LeeSahai2003} were aiming at the flow structure of a single narrow jet (or bullets),
while we are aiming at explaining the presence of two close jets (or bullets), i.e., twin-jets.

\cite{Balicketal2013} imaged CRL618 (PN G166.4–06.5) on few occasions from 1998 to 2010, and followed its evolution.
The pre-PN CRL618 contains two close narrow main lobes (finger-shaped outflows) on opposite sides of the center \citep{TrammellGoodrich2002}.
The lobes themselves show substructures, that, as we see later, are observed also in IRAS~4A2.
By jets we will refer also to very brief jets that are termed bullets.

The opposite twin-jets in CRL618 are not exactly aligned with each other.
The projected angle between the east twin is $20^\circ$, while that of west twin is $12^\circ$.
We are not aiming at explaining the structure of the narrow lobes (for that see \citealt{LeeSahai2003, Leeetal2003, Leeetal2009, Balicketal2013}),
but rather we examine the presence of twin-jets.
\cite{Balicketal2013} find that brief ejection of bullets can account for the structure of CRL618.
In section \ref{sec:CRL618} we discuss the properties of the bullets.
In section \ref{sec:IRAS4A2} we compare the non-identical twin-jets structure of CRL618 with that of the YSO IRAS~4A2.
In IRAS~4A2 the image does not reveal the twin-jets, but rather we had to carefully examine the position-velocity (P-V) maps to identify
this structure.
In section \ref{sec:summary} we speculate on a simple model to account for the formation of twin-jets.

\section{THE BRIEF BULLETS EJECTION IN CRL618}
\label{sec:CRL618}

Based on the observations and numerical simulations of CRL618 as performed by \cite{Balicketal2013} and \cite{LeeSahai2003}
we can construct the following plausible set of properties for the four lobes (two opposite twin-lobes) observed with HST.
The age of the lobes is $\sim 100 \yr$, implying that the bullets were launched within a much shorter time period.
For our modelling this time will be very-brief, $\tau_b \simeq 0.1 \yr$.
\cite{Balicketal2013} simulated the evolution of a bullet of mass $2.4 \times 10^{-5} M_\odot$ and with a velocity of
$300 \km \s^{-1}$.
We take the typical bullet initial mass to be $M_b \sim 5 \times 10^{-5} M_\odot$,
and the total mass in the four jets (that have more than 4 bullets) to be in the range $M_{4j} \sim 2-5 \times 10^{-4} M_\odot$.
If the mass in the jets is $\sim 0.1$ of the mass that was accreted onto a companion, then the accreted mass and the
accretion rate are $M_{\rm acc} \sim 3 \times 10^{-3} M_\odot$ and $M_{\rm acc} \sim 0.03 M_\odot \yr^{-1}$, respectively.

The energy of the accretion process that can be channelled to jets (bullets) and radiation, assuming accretion
onto a main sequence star, is
$E_{\rm ILOT} \simeq 0.5 G M_\odot  M_{\rm acc} /R_\odot \simeq 5 \times 10^{46} \erg$.
Radiation can come from the accretion process and from the interaction of the bullets with their surroundings.
{{{ Some ingredients of the model were discussed in the past in other contexts.
\cite{SokerRappaport2000} considered the interaction between AGB winds and high-momentum jets that are blown simultaneously,
and \cite{Blackmanetal2001sec} pointed out that accretion power in PN progenitors can be relatively steady and powerful.  }}}

\cite{SokerKashi2012} suggested that some structures in some PNe are formed by a brief energetic event
lasting weeks to years, that can be observed as an intermediate luminosity optical transient (ILOT).
An event with an energy of ${\rm few} \times 10^{46} \erg$ lasting few months can be in principle classified as an ILOT
similar to some ILOTs suggested for PNe \citep{SokerKashi2012}.
However, if dust obscures the inner several$\times \AU$, such an event will not be detected as an ILOT.

It should be noted that we refer only to the four lobes studied by \cite{Balicketal2013} and \cite{LeeSahai2003}, and not to the
massive molecular fast outflow, $M_{\rm mol} \sim 0.1 M_\odot$ \citep{SanchezContrerasetal2004}
that is closer to the center \citep{Kastneretal2001, Coxetal2003, SanchezContrerasetal2004, Nakashimaetal2007, Leeetal2009, Leeetal2013, Tafoyaetal2013}.
We attribute the fast molecular outflow closer to the center to a more vigourous recent binary interaction.
Namely, the same companion returned again in its eccentric orbit, but after losing angular momentum
and energy. It hence dived deeper into the primary AGB star.

An ILOT event was simulated recently by \cite{AkashiSoker2013}, who launched a wide massive outflow into a dense
AGB wind close to the center. The jets-shell interaction is within a distance of $r_I \sim 10 \AU$.
The mass in each jet was $\sim 0.01 M_\odot$ and its initial velocity was set to $1000 \km \s^{-1}$.
In this regime the radiative cooling of the gas is set by photon diffusion time, and that flow was found to be cooling slowly and
be unstable. The bullets proposed for CRL618, on the other hand, seem to cool rapidly.

Within months of a periastron passage the companion launched two opposite twin-jets (twin-bullets), each with a velocity of
$v_b \simeq 500 \km \s^{-1}$. Each bullet reached a distance of $\sim 10 \AU$ within the interaction time.
We take the typical interaction of jets with their dense AGB wind to be at a distance of $r_I= 3 \AU$.
The ratio of the photon diffusion time out of the interaction region of a bullet to the flow time $r_I/v_f$ does not depend on the
opening angle of the jet and is given by \citep{AkashiSoker2013}
\begin{eqnarray}
\frac {\tau_{\rm diff}} {t_{\rm f}} =
\frac{M_b \kappa}{ 4 r_I^2 } \frac{v_b}{c}
\simeq 0.007
\left( \frac {M_b}{5 \times 10^{-5} M_\odot} \right)
\left( \frac {v_b}{500 \km \s^{-1}} \right)
\left( \frac {r_I}{3 \AU} \right)^{-2},
\label{eq:tfdiff}
\end{eqnarray}
where $\kappa=0.34$ is the opacity.
This ratio implies that the bullets rapidly cool and maintain the narrow flow, unlike the wide jets simulated by \cite{AkashiSoker2013}.
The few bullets that can be formed along each of the four lobes can result from the stochastic nature of the launching process, or from interaction
with the surrounding gas.

\section{A TWIN-JET STRUCTURE IN NGC~1333~IRAS~4A2}
\label{sec:IRAS4A2}

\begin{figure}
\begin{center}
\includegraphics[scale=0.5]{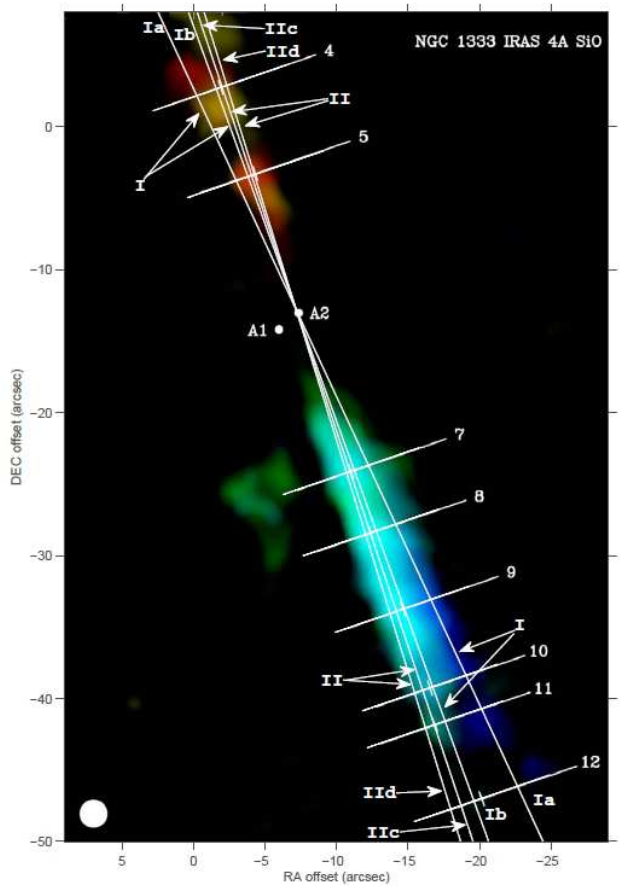}
        \caption{The color composite image of the NGC 1333 IRAS 4A2 bipolar jet in the SiO
$v=0\;J=1\rightarrow0$ line from \citet{Choi2011a}.
The white lines across the jets that are labelled by numbers, the `cuts', are from the original figure of \citet{Choi2011a},
while the four lines along the two opposite jets are our addition.
These four lines mark the two opposite twin-jets ($I$ and $II$), each twin composed of two components (marked by letters).       }
\label{fig:jets}
\end{center}
\end{figure}

The jets of the YSO IRAS~4A2 are presented in Figure \ref{fig:jets}.
Although the properties of the jets are not exactly as those of CRL618, we here find similarities in the twin-jets structure.

\begin{figure}
\begin{center}
\includegraphics[scale=0.5]{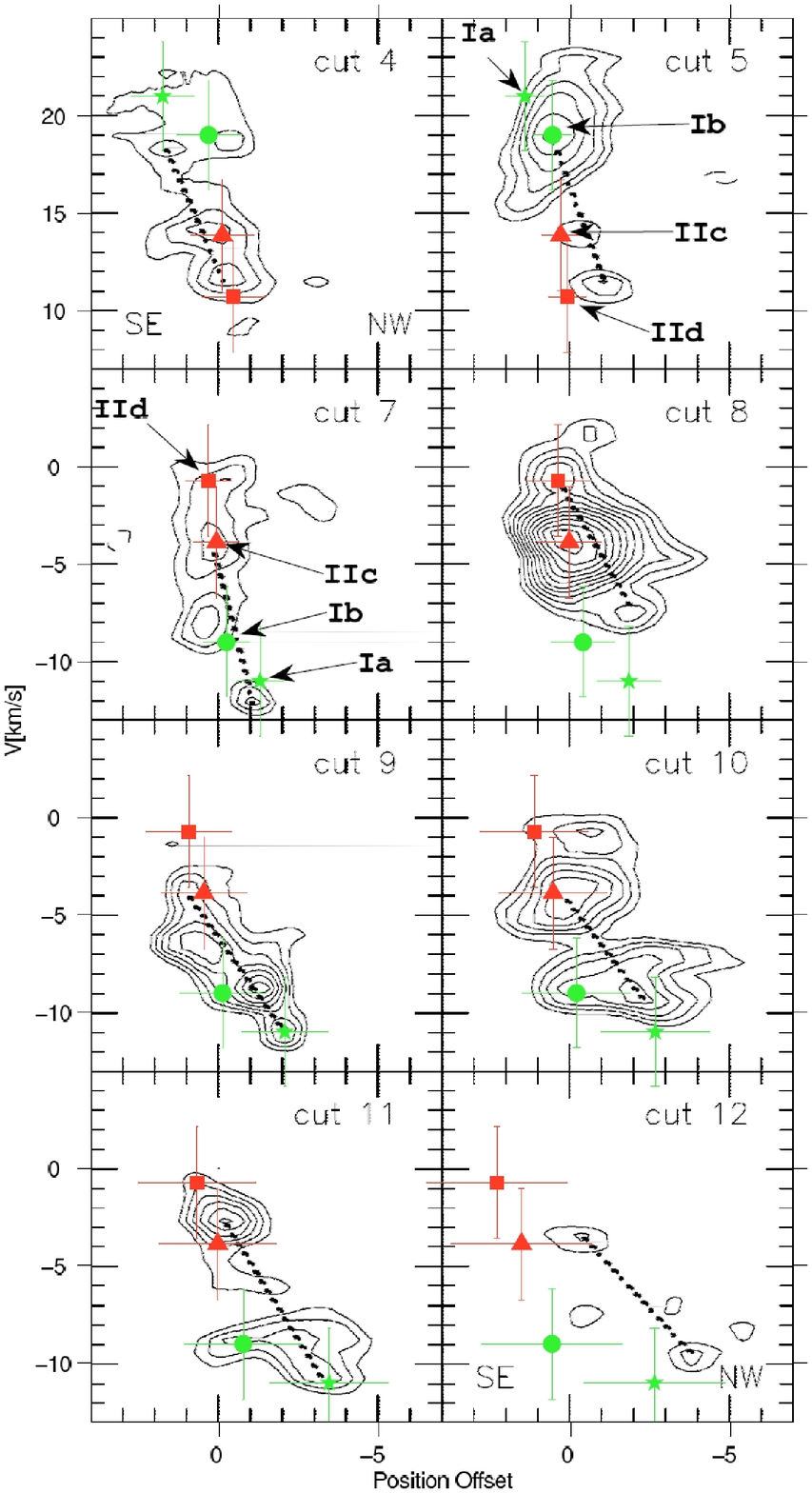}
        \caption{The position-velocity (P-V) diagrams from \citet{Choi2011a}.
        The colored crosses are our addition.
        The center of each cross represents the Doppler shift and position of a twin-jet's component along the appropriate cut.
        Each of the four symbols of the crosses represents one component, as marked on two of the panels.
        }
   \label{fig:PV1}
\end{center}
\end{figure}

\citet{Choi2011a} used the Very Large Array to observe the bipolar jets of IRAS~4A2 in the SiO $v = 0 \; J = 1\rightarrow 0$ line.
They investigated the kinematics of the two opposite SiO jets by placing
slits perpendicular to the jets' axis; these were termed cuts.
The jets and the positions of the cuts (marked by numbers) are shown in Figure \ref{fig:jets}.
\citet{Choi2011a} produced P-V diagrams for each cut, where
they positioned the zero displacement of each cut at the emission peak according to \cite{Choi2005}.
The P-V diagrams of the perpendicular slits are presented in Figure \ref{fig:PV1} here, taken from Figure 2 of
\citet{Choi2011a}, where the colored crosses are our addition which we discuss below.


The P-V diagram of each cut shows a multi-peak, three to four, structure, like the
clear separated structures in cuts 10 and 11.
The central peak of cut 8 is the brightest, and might be the overlap of two peaks.

We try to fit the peaks with twin-jets, where each twin jet is composed of two `components'.
Here a twin-jet refers to both northern and southern parts. Namely, a twin-jet has two opposite parts relative to the central star.
The same holds for a 'component'.
We model each of the four components as a straight line of gas on the two opposite sides of the star, as drawn on Figure \ref{fig:jets}.
The axis of each component is parametrized by two angles:
the inclination relative to the plane of the sky, $\theta$, and the angle on the plane of the sky relative to component~Ia,
$\phi$.
For $\theta > 0$ the northern part points away from us, i.e., red-shifted.
The values of $\theta$ and $\phi$ are given in Table \ref{tab:Table1}.
Based on \citet{Choi2006} we take all components to have the same radial expansion velocity of $v_j = 71 \km \s^{-1}$.
We compute the P-V diagram of the four components and draw them as colored crosses on Figure \ref{fig:PV1},
where the velocity and position are at the center of the cross.
The crosses themselves represent our crude estimations of the errors in the derived values of $\theta$ and $\phi$. From
our trials of different fittings we crudely estimate the errors to be of few degrees.
\begin{table}
\centering
  \caption{The angles of each component}
    \begin{tabular}{rrrr}
          &              &  \\
    \hline
\multicolumn{1}{c}{Sub-jet}& \multicolumn{1}{c}{Component}& \multicolumn{1}{c}{$\theta$}  & \multicolumn{1}{c}{$\phi$}    \\
 \hline
 \multicolumn{1}{c}{I} & \multicolumn{1}{c}{Ia}& \multicolumn{1}{c}{$12.8^\circ$}      & \multicolumn{1}{c}{$0^\circ$}   \\
 \multicolumn{1}{c}{I} & \multicolumn{1}{c}{Ib}& \multicolumn{1}{c}{$11.2^\circ$}      & \multicolumn{1}{c}{$5.0^\circ$} \\
 \multicolumn{1}{c}{II}& \multicolumn{1}{c}{IIc}& \multicolumn{1}{c}{$4.6^\circ$}      & \multicolumn{1}{c}{$6.5^\circ$} \\
 \multicolumn{1}{c}{II}& \multicolumn{1}{c}{IId}& \multicolumn{1}{c}{$7.1^\circ$}      & \multicolumn{1}{c}{$7.7^\circ$} \\
 \hline
   \end{tabular}
  \label{tab:Table1}
\end{table}

The four components (Ia, Ib, IIc, IId) do not perfectly fit the observed P-V diagram as can be seen in Figure \ref{fig:PV1}.
First, there is a substantial interaction of the jets with the ambient gas.
This is clearly evident from the sharp bend of the northern jet which occurs further out (\citealt{Choi2006}; not seen here).
Other regions also show some interaction with the surroundings, as evident from the wiggling boundaries of the jets.
Second, this interaction or jittering of the jets' source might cause random change in the expansion directions of the
different components.

\begin{figure}
\begin{center}
\includegraphics[scale=0.5]{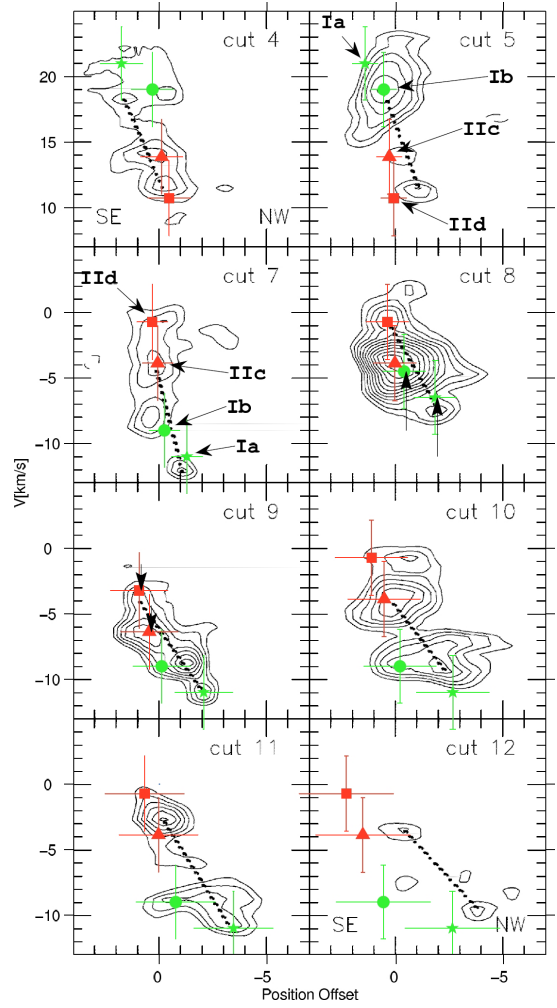}
        \caption{Like Figure \ref{fig:PV1}, but with a displacement of twin-jet I in cut 8 and of twin-jet II in cut 9, as marked
        by the vertical black arrows.       }
   \label{fig:PV2}
\end{center}
\end{figure}

We notice that we can substantially improve the fitting by displacing components Ia and Ib together in cut 8,
and components IIc and IId in cut 9.
The displacement is $\Delta \theta = -3.7^\circ$, namely redward in the southern jet, for components Ia and Ib in cut 8,
and $\Delta \theta = 2.0^\circ$, namely blueward in the southern jet,  for components Ic and Id in cut 9.
The P-V diagram with these displacements is presented in Figure \ref{fig:PV2}.
The displacement in cut 8 explains why the peak there is as twice as high as in the other cuts.
As the simultaneous displacement of components Ia and Ib accounts quite well for the P-V diagram in cut 8, we identify
them as a single twin-jet I.
As well, we identify components IIc+IId as twin-jet II.
The displacement in cuts 8 and 9 we apply here seems ad-hoc and speculative. But noticing
that the angels between the east and west twin-jets of the PN CRL618 are different, this turns out to be a very plausible explanation.

Our main conclusions from the simple fitting, although not perfect, of the P-V diagram of IRAS~4A2 are as follows.
\begin{enumerate}
\item The two opposite jets of IRAS~4A2 are actually composed of two bipolar twin-jets, reminding us of the narrow lobes in CRL618.
Each of the bipolar twin-jets, I and II,  is composed of two components, Ia and Ib, and IIa and IIb, respectively.
\item We can make the fitting with two twin-jets that have the same radial velocity, and with a very small angle between them.
These suggest a common launching site.
\item There is no sign of precession. The twin-jets and the two components composing each of them maintain an average
constant direction. We note that \cite{Balicketal2013} writes: ``The fingers of CRL618 are straight and show
no sign of a precessing or varying jet launcher.''
\item The structure of four-components and two twin-jets exists on both sides of the star (north and south jets).
This shows that the twin-jets structure does not result from a random collision and stochastic interaction
with the surrounding gas.
\item The relatively large displacement of one twin-jet we identify in cuts 8 and 9, according to our interpretation,
can result from either change in launching direction, as is suggested by the structure of the twin-jets
in CRL618, or from interaction with the surrounding gas.
\item No extra component of jet rotation is required in our interpretation.
\end{enumerate}

\section{DISCUSSION AND SUMMARY}
\label{sec:summary}

Our interpretation of the P-V diagram of IRAS~4A2 of two twin-jets structure requires that the launching process
has a preferred direction that does not change with time, at least not for the time period span by the jets, of $\sim 500 \yr$
for IRAS~4A2.
This excludes the possibility that the preferred direction is the direction toward a companion in a circular orbit.
For a slowly varying direction toward the companion the orbital period must be thousands of years or more.
Such a companion will be at a too large distance to influence the launching process of the jets.

However, if the orbit is eccentric the semi-major axis of the orbit defines a preferred direction that can maintain
a constant direction for a very long time.
We proposed therefore, that the launching process of the jets in IRAS~4A2 is substantially influenced
by the presence of a companion on a highly eccentric orbit.

\cite{Ardila2005} present the interaction of a disk with a close binary star on a parabolic orbit.
This passage creates two large-scale spiral arms in the disk ($m=2$ mode).
The spiral arms extend inward to a distance of $\sim 0.1 r_p$, where $r_p$ is the periastron distance,
and preserve a more or less constant direction for most of the time.
The dense-spiral arms structure in the gas component of the disk lasts for a time of $\sim 0.5 P_p$, where $P_p$ is the orbital period of a
circular orbit of radius $r_p$.
The two spiral arms can lead to the formation of the two sub-jets, i.e. twin-jets, that we identify in the outflow from IRAS~4A2.
{{{ Spiral arms in the inner part of a Keplerian disk, but in the $m=1$ mode and with no fix orientation, are also discussed for some
models of supergiant Be binary star systems (e.g., \citealt{Okazaki1997}). }}}

We can estimate plausible parameters for the orbit in IRAS~4A2.
Taking the IRAS~4A2 jets' velocity of $70 \km \s^{-1}$ to be the escape speed from their launching radius $r_L$,
we find $r_L \simeq 6 R_\odot$, where the mass of the primary star is $M_\ast=0.08 M_\odot$ \citep{Choi2011b}.
Let the companion periastron be then at $r_p \simeq 3-10 r_L \simeq 20-60 R_\odot$, and the apastron be at $\xi \gg 1$ times this distance.
The orbital time of the binary system (assuming the companion to have a mass of $M_2 << M_1=0.08M_\odot$) is
\begin{equation}
P_{\rm orb} = 1.1
\left( \frac{M_\ast}{0.08M_\odot} \right)^{-1/2}
\left( \frac{r_p}{50 R_\odot} \right)^{3/2}
\left( \frac{1+\xi}{4} \right)^{3/2} \yr.
\label{eq:Porb}
\end{equation}
Each jets' launching episode lasts for a few months.
Over the hundreds of years life time of the jets, the many launching episodes will be smeared to continuous jets.
While the system is near apastron, the disk near the primary star can rebuild itself.

One prediction of the model we proposed is that a careful monitoring of IRAS~4A2 will reveal a periodic activity with a period time
between few months and several years.
Observations should be ``lucky'' enough to catch the system near a periastron passage, when massive launching takes place.
The companion is a brown dwarf or a massive planet.

We cannot elaborate on how the two spiral arms in the disk lead to the formation of two sub-jets.
Simply, there is no agreed-upon model for the launching of jets that we can use to derive the process.
As for the two components of each twin-jet (sub-jet; e.g, Ia and Ib of twin-jet I), numerical simulations of the gravitational interaction of the companion with the
disk around the primary star should be carried out, probably in three-dimensions. Such simulations can reveal the presence of
higher harmonics in the accretion disk in addition to the spiral arms, as well as the development of instabilities. But again,
the way the structure in the disk influences the jets formation is a more complicated process.

Based on this interpretation, we also speculate that the central star of CRL618 has a main sequence binary companion on an eccentric orbit.
\cite{Leeetal2003} commented that the multipolar ejections at different directions in CRL618 is probably due to the presence of a binary companion.
We go further and attribute the interaction to an eccentric orbit.
The periastron distance is about the radius of an AGB star, such that high-rate mass transfer process could have taken place during the several
weeks of the periastron passage. This event could have been classified as an ILOT event about a century ago.
The orbital period can be from few years to tens of years (if the orbit is highly eccentric, as in $\eta$ Carinae).
In CRL618 it is the companion that launched the twin-jets after accreting the mass.
In our scenario the AGB star induced the double-arm spiral structure in the accretion disk that led to the formation of the twin-jets.
A companion on an eccentric orbit can account also for the departure from axi-symmetry of CRL618 (i.e., the two twin-jets are not exactly opposite to each other).

\cite{Balicketal2013} proposed that the several bullets on each side of CRL618 result from a spray of bullets
formed from instabilities in a rapid mass ejection, such as in the process proposed by \cite{Blackmanetal2001} and \cite{Mattetal2006}.
We note though that a model based only on instabilities will not explain a constant direction for each twin-jet over a long time,
as is the case with IRAS~4A2 that has long lasting jets.
This is the reason we propose that the preferred direction determined by the two twin-jets is a result of a binary eccentric orbit.

{{{ We thank an anonymous referee for useful comments.
This research was supported by the Asher Fund for Space Research
at the Technion, and the US-Israel Binational Science Foundation. }}}

\end{document}